\documentclass[12pt]{article}
\usepackage{authblk}
\usepackage{amsfonts,amssymb,amsmath}
\usepackage{tikz}
\usepackage{caption} 
\usepackage[marginparwidth=2.5cm]{geometry}
\usepackage{marginnote}

\newcommand{\Hom}{{\rm Hom}}

\newcommand{\cG}{{\mathbb G}}
\newcommand{\cC}{{\mathcal C}}
\newcommand{\ZZ}{{\mathbb Z}}
\newcommand{\RR}{{\mathbb R}}

\newcommand{\eqnref}[1]{Eq.~\eqref{#1}}
\newcommand{\figref}[1]{Fig.~\ref{#1}}

\newcommand{\secref}[1]{Sec.~\ref{#1}}

\newtheorem{question}{Question}
\newtheorem{theorem}{Theorem}

\usetikzlibrary{decorations.markings}
\usetikzlibrary{positioning}
\usetikzlibrary{shadings}
\RequirePackage[normalem]{ulem} 
\RequirePackage{color}\definecolor{RED}{rgb}{1,0,0}\definecolor{BLUE}{rgb}{0,0,1} 

\usetikzlibrary{decorations.markings}
\usetikzlibrary{positioning}
\usetikzlibrary{shadings}

\begin{document}

\title{Higher SPT's and a generalization of anomaly in-flow}
\author[1]{Ryan Thorngren}
\affil[1]{Department of Mathematics, University of California, Berkeley, CA}

\author[2]{Curt von Keyserlingk}
\affil[2]{ Princeton Center for Theoretical Science, Princeton University, Princeton, New Jersey 08544, USA}

\date{\today}
\maketitle

\begin{abstract}
Symmetry protected topological (SPT) phases of bosons in $d$ spatial dimensions have been characterized by the action of the protecting global symmetry $G$ on their boundary. The symmetry acts on the boundary in a way that would be impossible to realize in a purely $d-1$ dimensional system i.e., without the bulk. This is often formalized by saying the $G$ symmetry is anomalous when the boundary theory is gauged. Simultaneously gauging the symmetry on the boundary and in the bulk yields a gauge-invariant composite system. One says there is an anomaly in-flow from the boundary to the bulk. Recently it has been appreciated that some anomalies are too severe to be regulated by an SPT bulk. These do not satisfy anomaly in-flow in the traditional sense. However, we show that these anomalous systems can be regulated as the symmetric boundaries of the newly discovered higher SPT phases. These higher SPT's are protected by a symmetry $d$-group, a higher categorical version of a group which can charge not just particles but also strings, volumes, etc. This structure emerges naturally from computation of the anomaly. One can interpret these severe anomalies as producing an emergent higher-form gauge field when $G$ is gauged, and this gauge field participates in the bulk topological action.
\end{abstract}

\section{Introduction}

In this paper, we study 't Hooft anomalies for finite symmetry groups $G$. This is the situation where a field theory has a global $G$ symmetry which cannot be gauged. For simplicity, especially when we talk about gauging, we only consider internal symmetries (no time reversal, fermion parity, etc).  By gauging we mean we couple the field theory to a flat $G$ gauge field and attempt to preserve $G$ gauge invariance by introducing local counterterms involving the original fields and the $G$ gauge field. When it is not possible to assemble such counterterms, we say the system has an 't Hooft anomaly for the symmetry $G$.

In such situations, the variation of the $d$-dimensional action, while not cancellable by local counterterms, can often be cancelled by the boundary variation of a $d+1$-dimensional local term for the $G$ gauge field. In such a situation we say $G$ satisfies {\it anomaly in-flow} and the $d+1$-dimensional TQFT of the $G$ gauge field is a sort of universality class of the 't Hooft anomaly, preserved under RG flow and small perturbations. A basic question is
\begin{question} 
Do all symmetries satisfy anomaly in-flow?
\end{question}
In \cite{KT} this question was answered negatively in the field theory context. The authors demonstrated a 2+1d Dijkgraaf-Witten theory with an anomalous $G$ symmetry too severe to be cancelled by any 3+1d theory of just the $G$ gauge field. This is related to the $H^3$ obstruction to finding a $G$-crossed braided extension of a modular tensor category that was studied in a mathematical context in \cite{ENO} and discussed in a condensed matter context in \cite{CLV,BBCW} (see also \cite{Fidkowskinew}).

In this paper however, we return to the example studied in \cite{KT} and show that if we let the $G$ gauge field in 3+1d be the 1-form part of a 2-gauge field (which includes also a 2-form field) then there is a local term whose boundary variation cancels the anomaly. In fact we show
\begin{theorem}
All anomalies of Dijkgraaf-Witten theories satisfy anomaly in-flow after extending $G$ to a higher symmetry.
\end{theorem}
We will show this in the field theory after $G$ has been gauged using the Serre spectral sequence, which naturally filters the anomalies by their severity (and also by how ``higher" the symmetry must be extended).

Further, in 2+1d, where there is a concrete understanding of a fairly general class of TQFT's, we show 
\begin{theorem}
Any $G$ symmetry of a modular 2+1d TQFT satisfies anomaly in-flow after extending $G$ to a 2-group with $\Pi_2$ equal to the group of abelian anyons in the TQFT.
\end{theorem}
Along the way, we relate concretely the $H^3$ obstruction we observed in field theory to the $H^3$ obstruction found in \cite{ENO}. Our understanding of this obstruction is that it signals an emergent 1-form gauge symmetry that comes along with gauging $G$. Indeed, the emergent codimension 2 symmetry defects that proliferate are composed of ``associator bubbles" of $G$ domain walls, where the $H^3$ cocycle determines the charge of the bubble. 

These sorts of questions have received much recent attention from condensed matter physicists since the discovery of symmetry protected topological (SPT) phases. These are gapped systems with $G$ symmetry which, after gauging $G$ and passing to the IR TQFT, we get some non-trivial topological term. If the original system has a symmetric boundary, then $G$ has to have an 't Hooft anomaly on the boundary to cancel the variation of the bulk topological term. The analog of question 1 in this ``ungauged" case is
\begin{question}
Are all $G$-symmetric systems realizable on the boundary of a $G$ SPT phase with an on-site group action?
\end{question}
It is widely believed that $G$ acting on-site in some theory with the same IR fixed point as our system implies the $G$ symmetry is non-anomalous. Given this, since the answer to question 1 is no, the answer to question 2 is also no.

However, in \cite{KT2} the notion of SPT phase was generalized to higher symmetries, which in a lattice model act not just on-site in the non-anomalous case but on vertices, edges, faces, etc. in a controlled way and whose symmetries form not a group but a $d$-group in $d$ dimensional spacetime. We call such a symmetry action ultralocal. In this sense, we show
\begin{theorem}
Any $G$-symmetric Dijkgraaf-Witten theory is realizable on the boundary of a higher SPT phase with ultralocal symmetry action.
\end{theorem}

All our conclusions can be re-interpreted in terms of anyon condensation using the observation that any abelian anyon acts as the generator of a 1-form symmetry. Gauging the 1-form symmetry means proliferating symmetry defects means condensing these anyons. Thus, another way to understand the $H^3$ anomaly is that certain anyons become condensed when the anomalous 0-form symmetry is gauged. The charge of the condensate is determined by the background 1-form gauge field via the class in $H^3$. The remaining anomaly in $H^4$ is an obstruction to forming this condensate in a purely 2+1d system, but the condensate can be consistently formed (as far as TQFT rules are concerned) on the boundary of a 2-SPT in 3+1d.

\section{The Motivating Example: Field Theory}
\label{Background}

\subsection{Projective Symmetry Action}

Following \cite{KT}, we consider a 2+1d Dijkgraaf-Witten theory with gauge group $G_{gauge} = \ZZ/3$ and gauge field $a$. We consider this theory with the level 1 action
$$
S_{3d} = \frac{1}{3} \int \tilde a \cup \frac{\delta \tilde a}{3}.
$$
See appendix A for an introduction to our field theory notation.

In \cite{KT}, the authors considered gauging a global $G_{global} = \ZZ/3 \times \ZZ/3$ symmetry of this theory. This introduces a new $\ZZ/3 \times \ZZ/3$ gauge field $A = (A^1,A^2)$ and turns the flatness constraint for $a$ into
\begin{equation}\label{flatness}
\delta a = c(A) \mod 3,
\end{equation}
where $c$ is a group cocycle classifying a group extension of $G_{global}$ by $G_{gauge}$. We call the resulting extended group the total group of symmetries $G_{total}$. We henceforth use the extension $c(A^1,A^2) = A^1 \cup A^2$. 

This cocycle determines how $G_{global}$ acts projectively on Wilson lines. Indeed, the constraint \eqref{flatness} implies that in a non-trivial (but necessarily flat) $A$ background, the electric charge $q$ Wilson line
\begin{equation}\label{WilsonLine}
\exp{2\pi iq/3 \int_\gamma a}
\end{equation}
can change by a phase if the worldline $\gamma$ is deformed. In particular, if the worldline bounds a disc $\gamma = \partial D$ then the above equals
$$
\exp(2\pi iq/3 \int_D c(A)).
$$

\begin{figure}[t]
\centering
\begin{tikzpicture}
\node (pic) {\includegraphics[width=0.4\textwidth]{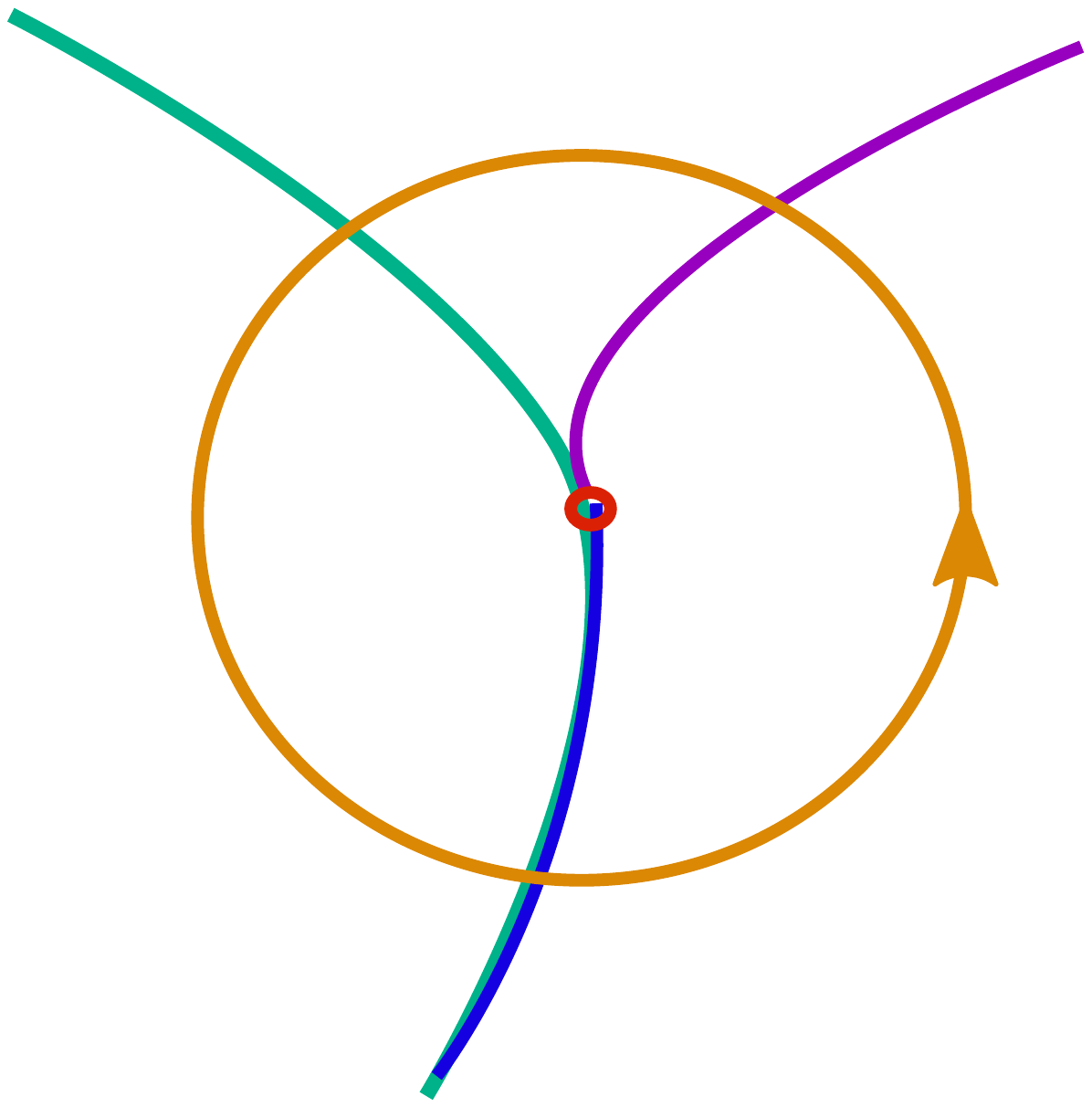}};
\node at (-2,2) {$y$};
\node at (1,2.3) {$x$};
\node at (-1,-1) {$yx$};
\node at (1.3,0) {$q$};
\end{tikzpicture}
\captionof{figure}{An electric charge encircles a codimension 2 defect where two domain walls $x, y$ fuse to a third. \label{zipper}}
\end{figure}

We can think of a configuration of the background $A$ as a network of $G_{global}$ domain walls with codimension 2 ``zippers" where pairs of domain walls fuse. Then the integral in the exponent can receive a contribution every time $D$ intersects one of these zippers. In particular, consider a zipper junction where a $x=(x^1,x^2)$ and $y=(y^1,y^2)$ domain wall meet, $x,y \in G_{global} = \ZZ/3 \times \ZZ/3$. Consider $\gamma$ to be a small loop linking this zipper (see \figref{zipper}). The cocycle $c$ defines an element $c(x,y) = x^1 y^2 \in G_{gauge} = \ZZ/3$, so that a charge $q$ test particle accrues a net phase
\[ \exp(2\pi i q x^1 y^2 /3),\]
 on braiding around the zipper (along the orange path in \figref{zipper}).

That the test particle acquires a phase on encircling a zipper is equivalent to the statement that the charge $q$ Wilson line transforms projectively under the $G_{global}$ symmetry. More precisely, we should view the charge $q$ Wilson lines as being the world-lines of particles whose internal Hilbert space carries a projective representation of  $G_{global}$. This in turn, is equivalent to the statement that these particles carry an honest faithful $G_{total}$ representation, in a manner we now describe. Recall that $G_{total}$ is obtained through an extension
\begin{equation}\label{eq:extension}
G_{gauge} \xrightarrow[j]{} G_{total} \xrightarrow[\pi]{} G_{global}.
\end{equation}
We will define the maps $\pi,j$ explicitly below. To describe the fundamental $G_{total}$ representation carried by the Wilson lines, we embed $G_{total}$ into the group of $3\times3$ unitary matrices $U(3)$ with generators
\[ X = \left( \begin{array}{ccc}
0 & 1 & 0 \\
0 & 0 & 1 \\
1 & 0 & 0 \end{array} \right)\]
\[ Y = \left( \begin{array}{ccc}
1 & 0 & 0 \\
0 & \exp(2\pi i/3) & 0 \\
0 & 0 & \exp(4 \pi i/3) \end{array} \right).\]
These satisfy
\[ XYX^{-1}Y^{-1}= \left( \begin{array}{ccc}
\exp(2\pi i/3) & 0 & 0 \\
0 & \exp(2\pi i/3) & 0 \\
0 & 0 & \exp(2 \pi i/3) \end{array} \right) =: Z,\]
defining a central element $Z$ (i.e., one which commutes with everything). This relation along with $X^3 = Y^3 = Z^3 = 1$ characterize $G_{total}$. Now we define the maps named above.
\begin{itemize}
\item The map $j:G_{gauge} = \ZZ/3 \to G_{total}$ sends the generator $1 \in \ZZ/3$ to $Z$.

\item The map $\pi:G_{total} \to G_{global} = \ZZ/3 \times \ZZ/3$ sends
\[ X \mapsto (1,0)\]
\[ Y \mapsto (0,1).\]

\end{itemize}
One checks that these maps make the sequence \eqref{eq:extension} a central extension. It will be necessary to choose once and for all a section $s:G_{global} \to G_{total}$. That is, we want a set map satisfying $\pi \circ s = id$. This section will be used to define the projective $G_{global}$ action. It will not be possible to choose $s$ to be a group homomorphism (one says \eqref{eq:extension} is not splittable). We choose
\[s(m,n) = X^nY^m.\]

The fundamental ($q=1$) Wilson lines in this theory should be through of as the world-lines of particles whose internal Hilbert space transforms in this three dimensional representation just described. We can associate a wavefunction $|\psi\rangle$ with the internal Hilbert space of the particle. $|\psi\rangle$ is an automatically an eigenvector of central element $Z$, which generates $G_{gauge}$, with eigenvalue $\exp(2\pi i/3)$, so these are all charge $q=1$ from the perspective of $G_{gauge}$. A global symmetry transformation $(a,b)\in G_{global}$ acts by
\[ |\psi\rangle \mapsto s(a,b)|\psi\rangle \]
using the section just described. Thus, a particle prepared in state $|\psi\rangle$ which travels over the zipper ends up in state
\[ s(g_2)s(g_1)|\psi\rangle, \]
while one traveling below ends up in state
\[ s(g_2g_1) |\psi\rangle.\]
These differ by the operator $Z^{- g_1^1 g_2^2}$, which acts by a phase. This phase equals \eqref{WilsonLine} when $q=1$.

In the category language we have essentially tripled the collection of electric quasiparticles by introducing these extra internal degrees of freedom on the charges. The global symmetry generator $X$ permutes these three copies, while the global symmetry operator $Y$ acts by a different $\ZZ/3$ charge in each copy. The relation $[X,Y] = 0$ holds only up to the gauge transformation $Z$.

\subsection{Anomaly In-Flow}\label{ss:anomalyinflow}

The flatness condition for $a$ was modified by gauging $G_{global}$, and as a result the bare action $S_{3d}$ is no longer $G_{gauge}$ invariant. Consider a shift $\tilde a \mapsto \tilde a + 3x$. This causes a shift
\[ S_{3d} \mapsto S_{3d} + \frac{1}{3}\int \tilde a\cup \delta x,\]
modulo integer terms. Previously this term would have disappeared using integration by parts, but that is no longer the case because of the modified flatness constraint, which gives instead a term
\[ -\frac{1}{3} \int \tilde c(A) \cup x.\]
This term must be cancelled by redefining
\[ S'_{3d} = S_{3d} - \frac{1}{9} \int \tilde c (A) \cup \tilde a.\]
Writing $\delta \tilde a = \tilde c(A) + 3 \alpha$, we can rewrite the action
\[ S'_{3d} = \frac{1}{3} \int \tilde a \cup \alpha - \frac{1}{9} \int \tilde c(A) \cup \tilde a.\]
Now consider a $\ZZ/3$ gauge transformation $a \mapsto a + \delta f$, but in terms of the integer lifts $\tilde a \mapsto \tilde a + \delta \tilde f \mod 3$. Since we have already shown invariance under shifts of $\tilde a$, we need only show invariance under the lifted gauge transformation $\tilde a \mapsto \tilde a + \delta \tilde f$. Under this variation the action transforms
\[ S'_{3d} \mapsto S'_{3d} + \int \frac{1}{3} \delta \tilde f \cup \alpha - \frac{1}{9} \tilde c(A) \cup \delta \tilde f.\]
We can rewrite this variation using $\delta \alpha = \delta \tilde c(A)/3$
\[ - \int \frac{1}{9} \tilde f \cup \delta \tilde c (A) + \frac{1}{9} \tilde c(A) \cup \delta \tilde f.\]
The quantities $\tilde c(A)$ and $\delta \tilde f$ commute up to exact terms modulo 3. However, they appear with the denominator 9, so when we bring them past each other, there is a cup-1 correction term to deal with by a shift
\[S'_{3d} \mapsto S'_{3d} - \frac{1}{3} \int \frac{\delta \tilde c (A)}{3}\cup_1 a\]
(see \cite{KT} for a more detailed discussion). Given this correction, which is safely ignored by the reader, the final variation is
\[ -\frac{1}{3} \int f \cup \frac{2\delta\tilde c(A)}{3}.\]
This term can be cancelled by a shift
\[ S'_{3d} \mapsto S'_{3d} + \frac{1}{3} \int a \cup B\]
iff 
\begin{equation}\label{Postnikov}
\delta B = 2\delta \tilde c(A)/3 = 2Bock(c(A)) =: \beta(A).
\end{equation}
The 3-cocycle $\beta(A)$ is called the Postnikov class. In \cite{KT} it was shown more systematically that this is the only possible local counterterm that can deal with this variation. In particular, $B$ should be only a function of $A$, so if the cohomology class $[2Bock( c )] \in H^3(BG_{global},G_{gauge}^*)$ is nonzero, there is no way to solve \eqref{Postnikov} for arbitrary configurations of $A$, so there is an anomaly.\footnote{Note that if we used $G_{gauge} = \ZZ/2$, there would be no anomaly, since this variation would be an integer.}

It was argued in \cite{KT} that this anomaly cannot be cancelled by placing the system on a $G_{global}$ SPT boundary. To see this, first note that our action is
\[ S''_{3d} = \int \frac{1}{3} \tilde{a} \cup \alpha - \frac{1}{9} \tilde c (A) \cup a - \frac{1}{3}Bock( c ) \cup_1 a.\]
Write the Lagrangian density $\omega(A,a)$. $G_{total}$ gauge invariance is equivalent to $\delta \omega = 0 \mod \ZZ$. As in \cite{KT}, one computes
\begin{equation}\label{SETaction}
\delta \omega = \frac{1}{3}a \cup \beta(A) + \tilde\Omega(A)
\end{equation}
for some 4-cochain $\tilde\Omega$. Anomaly in-flow in this situation means that by placing the 3d spacetime $X$ on the boundary of a 4d spacetime $Y$, the composite action
\[ \int_X \omega(A,a) - \int_Y \tilde\Omega(A) - \int_Y \frac{1}{3} a \beta(A)\]
is $G_{total}$ gauge invariant. However, when the third term is nonvanishing, the $G_{gauge}$ field $a$ must live on $Y$ rather than be constrained to the boundary. This gauge field is there even before gauging the $G_{global}$ symmetry, so the bulk state is not short range entangled before gauging; it is not an SPT. Rather we say it is a kind of symmetry enriched topological order, or SET for short.

However -- and this is the novelty of this paper -- instead of requiring $B$ be determined by $A$, we can let $B$ be a dynamical bulk 2-form satisfying equation \eqref{Postnikov} (and we constrain $A$ so that this constraint equation has solutions). In this case, it is always possible to cancel the anomaly using a 4d theory with action
\begin{equation}\label{s4d}
S_{4d} = \frac{1}{3} \int c(A) \cup c(A)/3  - c(A) \cup_1 \gamma_c (A) - B \cup c(A).
\end{equation}
Note that this action does not involve $a$! Neither, as in previous cases \cite{KT,ChenFidkowski}, is this action an $H^{4}(BG_{global}, \mathbb{R}/ \mathbb{Z})$ cocycle obtained from gauging a traditional SPT. Instead, this action come from gauging a higher symmetry of a short range entangled state. Indeed, this action describes a generalized Dijkgraaf-Witten theory for the {\it 2-group} we call $\cG_{global}$ defined by $\Pi_1 = G_{global} = \ZZ/3 \times \ZZ/3$, $\Pi_2 = G_{gauge}^* = \ZZ/3$,\footnote{We write $G_{gauge}^*$ to denote the Pontryagin dual of $G_{gauge}$. That is, $G_{gauge}^* = \Hom(G_{gauge},U(1)) = H^1(BG_{gauge},U(1))$. For finite abelian groups, they are isomorphic, but the $H^3$ anomaly lives naturally in $H^3(BG_{global},G_{gauge}^*)$, so we prefer this notation. It is also important to distinguish the $BG_{gauge}^*$ action, which is a global symmetry, from the $G_{gauge}$ gauge symmetry. See \secref{DWgeneral} for more details.} and Postnikov invariant $2\gamma_c$. In \cite{KT2} it was argued that such a TQFT is the gauged effective field theory of a certain generalized SPT phase (called a 2-SPT) protected by a symmetry action of this 2-group. That is, while our anomaly is too severe to live on the boundary of an SPT phase, it can live on the boundary of a 2-SPT phase!

\subsection{A Simpler $S_{4d}$}\label{sec:simpler4d}

Let us pause to give a simplification of $S_{4d}$. Let $B$ be a $\ZZ/3$ valued cochain satisfying \eqref{Postnikov}. We consider
\[
C = \tilde c(A) + 3\widetilde{B}.
\]
If we pick a different integer lift of $B$, $\widetilde{B}$ shifts by a term divisible by $3$, so $C$ shifts by a term divisible by $9$. Similarly, if we shift $\tilde c(A)$ by $3x$ for some integer 2-cochain $x$, because of the constraint \eqref{Postnikov}, $\widetilde{B}$ shifts by $-x \mod 3$, so again $C$ is invariant mod $9$. Therefore, for any configuration of the $\cG_{global}$ gauge field, we can produce a configuration of a $B\ZZ/9$ 2-form gauge field. In terms of $C$, the 4d action is rewritten as
\begin{equation}\label{4daction}
S_{4d} = \frac{1}{9}\int C \cup C,
\end{equation}
which can be verified by expanding $C^2$ mod $9$ using the definition of $C$ above.

\section{Local and nonlocal lattice symmetry}

\subsection{2-SPT Bulk}\label{2sptbulk}

In the bulk Hilbert space, there is a basis with states $|\phi,\lambda\rangle$ labeled by
\begin{itemize}
\item Elements $(\phi^1_i,\phi^2_i) \in G_{global}=\ZZ/3\times \ZZ/3$ for each vertex $i$.
\item Elements $\lambda_{ij} \in G_{gauge}^* = \ZZ/3$ for each edge $ij$ (not required to satisfy a flatness condition).
\end{itemize}
These are the same data that parametrize $\cG_{global}$ gauge transformations for $B$ and $A$ in the gauged TQFT's above. That is, starting from a trivial configuration $A=0$, $B=0$, we generate a flat $\cG_{global}$ gauge field by
\[ A(\phi,\lambda) = \delta \phi\]
\begin{equation}\label{B}
B(\phi,\lambda) = \delta \lambda + \beta_1(\phi),
\end{equation}
where $\beta_1(\phi)$ is the first descendant of the Postnikov class $\beta$,
\[ \beta_1(\phi)_{ijk} = \big[\phi^1_i (\phi^2_j - \phi^2_i) + \phi^1_j (\phi^2_k - \phi^2_j) > \phi^1_i(\phi^2_k - \phi^2_i)\big]\]

The $\cG_{global}$ symmetries are the most general transformations of $\phi, \lambda$ preserving $A(\phi,\lambda), B(\phi,\lambda)$. There are 0-form and 1-form symmetries.
\begin{itemize}
\item A global 0-form symmetry is simply an element $(g^1,g^2)\in \ZZ/3 \times\ZZ/3$ and acts by
\[ \phi^j \mapsto \phi^j + g^j \]
\[ \lambda_{ij} \mapsto \lambda_{ij} - \beta_2(\phi,g)_{ij},\]
where
\[\beta_2(\phi,g)_{ij} = g^1 (\phi^2_i > \phi^2_j) - g^2 (\phi^1_i > \phi^1_j)\]
is the second descendant of the Postnikov class $\beta$ defined in \eqref{Postnikov}. In this formula, $(g > h)$ is $1$ if $g>h$, written with representatives in $[0,3)$, and $0$ otherwise (see Appendix A). The strange transformation rule for the edge variables is the fault of the non-trivial Postnikov class. Without it, the 0-form and 1-form symmetries would decouple. See \cite{KT2}. Notice that even a global symmetry transformation transforms the edge variables if $\phi$ is not constant.

\item A global 1-form symmetry is parametrized by a $\ZZ/3$ 1-cocycle $\eta_{ij}$ (flatness is the right condition for 1-form symmetries to be global) and acts by
\[ \phi^j \mapsto \phi^j \]
\[ \lambda_{ij} \mapsto \lambda_{ij} + \eta_{ij}.\]

\end{itemize}
Note that the 1-form transformations cannot be used to trivialize the 0-form transformation rule for $\lambda$ since $\delta \beta_2 \neq 0$, so $\beta_2$ is not a valid choice of the parameter $\eta$. The 0-form and 1-form symmetries are truly coupled. This is contrasted with the case where the Postnikov class $\beta$ is exact, when we can perform this separation. This is true in a superficially very similar set up using $G_{global} = \ZZ/2 \times \ZZ/2$, $G_{gauge} = \ZZ/2$ and $c(A_1,A_2) = A_1 \cup A_2$ \cite{KT,ChenFidkowski}. The algebraic structure formed by the $\cG_{global}$ transformations is a 2-group, rather than simply a pair of groups.

The fact that these symmetries preserve the induced gauge fields $A, B$ mean the state
\[ |\Omega\rangle = \sum_{\phi,\lambda} \exp\big(2\pi i\int_{CX} \Omega(A,B)\big) |\phi,\lambda\rangle\]
is $\cG_{global}$ invariant. Here $CX$ is the cone over the spatial slice $X$ and $\Omega$ is the 3+1d TQFT Lagrangian density \eqref{4daction}. In the case that $X$ is closed, we can rewrite the state
\begin{equation}\label{groundstate}
|\Omega\rangle = \sum_{\phi,\lambda} \exp(2 \pi i\int_X \Omega_1(\phi,\lambda)) |\phi,\lambda\rangle.
\end{equation}
This is the bulk 2-SPT groundstate. Here $\Omega_1$ is the first descendant of $\Omega$. It is defined up to boundary corrections by
\[ \delta \Omega_1(\phi,\lambda) = \Omega(A(\phi,\lambda),B(\phi,\lambda)).\]
One computes from \eqref{4daction}
\begin{equation}\label{Omega1}
\Omega_1(\phi,\lambda) = \frac{1}{9} (c_1(\phi) + 3\lambda) \delta (c_1(\phi) + 3\lambda),
\end{equation}
where $c_1$ is
\[c_1(\phi)_{ij} = \phi_i^1(\phi^2_j - \phi^2_i).\]
We need $\lambda$ in the expression for the ground state $|\Omega\rangle$ to be invariant under global 0-form transformations. Indeed, when $\phi \mapsto \phi+g$,
\[ \frac{1}{9}c_1(\phi)\delta c_1(\phi) \mapsto \frac{1}{9}c_1(\phi)\delta c_1(\phi) + \frac{1}{3} Bock(c_2(\phi,g)) \delta c_1(\phi).\]
This variation is precisely cancelled by $\lambda \mapsto \lambda - \beta_2(\phi,g)$ since $\beta_2(\phi,g) = Bock(c_2(\phi,g)) \mod 3$. Thus, a global 0-form transformation preserves $\Omega_1$ and therefore also $|\Omega\rangle$. One easily checks the same for 1-form transformations.

\subsection{Nonlocal boundary action for 2-SPT}

The expression \eqref{groundstate} allows us to determine the nonlocal symmetry action on a boundary state of the 2-SPT. See \cite{EN} for a clear exposition of these ideas in the ordinary SPT case.

In the case that $X$ has boundary, the phases in \eqref{groundstate} cannot be written as integrals of $\Omega(A,B)$ over the cone $CX$, since $\partial CX = X \cup C\partial X$. Therefore, the invariance of $A, B$ under $\cG_{global}$ transformations does not imply that \eqref{groundstate} is invariant when $\partial X \neq 0$. Indeed, one can find a second descendant $\Omega_2$ satisfying
\begin{equation}\label{Omega2}
\delta \Omega_2(\phi,\lambda,g,\eta) = \Omega_1(\phi+g,\lambda + \eta - \beta_2(\phi,g)) - \Omega_1(\phi,\lambda).
\end{equation}
This causes a shift
\[(g,\eta) |\Omega\rangle = \exp\left(-2 \pi i \int_{\partial X} \Omega_2(\hat\phi,\hat\lambda,g,\eta)\right)\ |\Omega\rangle,\]
where $\hat\phi, \hat\lambda$ are the operators
\[\hat\phi|\phi,\lambda\rangle = \phi|\phi,\lambda\rangle\]
\[\hat\lambda|\phi,\lambda\rangle = \lambda|\phi,\lambda\rangle.\]
These operators commute so quantization of these descendants is unambiguous up to an overall phase. Note this rederives that $|\Omega\rangle$ is $\cG_{global}$ invariant when $\partial X = 0$. To have a symmetric state, we need to introduce boundary physics that transforms by an opposite phase. In the notation of \cite{EN}, this is the operator $\mathcal{N}^{(1)}$.

Further descendants may be computed as well. For instance, we may ask if the symmetry acts projectively on the boundary. In other words, we compute the difference between applying $(g_1,\eta_1)$ followed by $(g_2,\eta_2)$ and just applying $(g_1+g_2,\eta_1+\eta_2)$ all at once:
\[\Omega_2(\phi,\lambda,g_1,\eta_1) + \Omega_2(\phi+g_1,\lambda + \beta_1(0;g_1) + \eta_1)\]
\[- \Omega_2(\phi+g_1+g_2,\lambda + \beta_1(0;g_1+g_2)+\eta_1 + \eta_2)\]
\[ = \delta\Omega_3(\phi,\lambda,g_1,\eta_1,g_2,\eta_2)\]
for some 1-cochain $\Omega_3$. Thus, since the spatial boundary is closed, we get an honest global symmetry action. However, we find that if we restrict the action to a region $U$ of the boundary which itself has boundary, only applying the symmetry action to its interior, then there is a necessary boundary correction
\[ \int_{\partial U} \Omega_3(\hat\phi,\hat\lambda,g_1,\eta_1,g_2,\eta_2).\]
In \cite{EN}, this corresponds to the operator $\mathcal{N}^{(2)}$. Likewise one can compute the last descendant $\Omega_4$, which corresponds to $\mathcal{N}^{(3)}$. Note that the last descendant can be thought of as an element in $H^4(B\cG_{global},U(1))$, which characterizes this 2-SPT phase.


\subsection{Very nonlocal boundary action for SET}

We can contrast this boundary symmetry action with the symmetry action where we consider the 0-form $G_{global}$ symmetry on its own, rather than part of the $\cG_{global}$ 2-symmetry--concretely, there is no $\lambda$ field (or $B$ field in the gauged version \eqnref{s4d}) in the bulk. There is only $\phi \in G_{global}$ and the $G_{gauge}$ field $a$. Using the descendant procedure of \cite{EN} and the previous section we will find that as a pure 0-form symmetry, the $G_{global}$ operators cannot be implemented by finite depth local circuits. Then in the next section we show how to remedy this using the new field $\lambda$ in the 2-SPT bulk.

Consider the boundary of the SET described by the TQFT \eqref{SETaction}
\[ S_{SET} = \int_{4d} \tilde\Omega(A) + \frac{1}{3} a\beta(A).\]
It is possible to compute descendants of $\tilde\Omega(A) + a\beta(A)$ as in the SPT case. They have the form
\[ \tilde\Omega_j(\phi,g_1,...,g_{j-1}) - \frac{1}{3}a\beta_j(\phi,g_1,...,g_{j-1}),\]
where $\beta_j$ is the ordinary descendant we have seen appearing already, while $\tilde\Omega_j$ is something more complicated since $\tilde\Omega$ itself is not closed. For example, we can write down the ground states
\begin{equation}\label{SETGS}
|\omega,[a]\rangle = \sum_{\phi,f} \exp \big(2\pi i \int_X \tilde\Omega_1(\phi) - \frac{1}{3}(a + \delta f) \beta_1(\phi)\big) \ |\phi, a+\delta f\rangle.
\end{equation}
Recall that $\beta(A)$ is exact in the present context. As a result, the term $\frac{1}{3}a \beta $ in the SET action gives no contribution on a manifold without boundary. As a result, this system has a degenerate ground state for different gauge equivalence classes of flat connections $a$, labeled here $[a]$. Note also that the $G_{gauge}$ transformations act by
\begin{equation}\label{latticegaugetrans}
f_{gauge}|\phi,a \rangle = \exp\big(- \frac{2\pi i}{3} \int_X f \delta\beta_1(\phi) \big) |\phi, a + \delta f\rangle.
\end{equation}
This can be read off from the $a \beta(\delta\phi)$ term in the action, which says that $\beta(\delta\phi)=\delta\beta_1(\phi)$ acts as a $G_{gauge}$ current. On a closed spatial slice, \eqref{SETGS} is gauge invariant since we can integrate by parts
\[ \int f \delta \beta_1(\phi) = - \int \delta f \beta_1(\phi).\]
When $X$ has boundary, this produces an extra phase
\[ f_{gauge}|\omega,[a]\rangle = \exp\big( \frac{2\pi i}{3}\int_{\partial X} f \beta_1(\hat\phi)\big) |\omega,[a]\rangle,\]
which, if we are to preserve gauge invariance, must either be cancelled by a gauge anomaly on the boundary or a $G_{global}$-breaking boundary condition that forces $\beta_1(\phi) = 0\mod 3$. We call the operator on the right hand side $\mathcal{O}^{gauge}_\partial(f)$. This operator can be restricted to an operator $\mathcal{O}^{gauge}_U(f)$ supported only in a region $U\subset \partial X$ by restricting the support of $f$ to lie in $U$.

We also get some phase when we apply a $G_{global}$ symmetry. Indeed,
\[g |\omega,[a]\rangle = \exp\big(2\pi i \int_{\partial X} \tilde \Omega_2(\hat\phi,g) - \frac{1}{3}\hat a \beta_2(\hat\phi,g)\big)|\omega,[a]\rangle.\]
Here $\hat a$ refers to the operator with $\hat a|a\rangle = a|a\rangle$. This phase must be cancelled by an opposite variation of the boundary state if the whole system is to be symmetric. We call the operator on the right hand side $\mathcal{O}^{global}_\partial(g)$. This operator can also be restricted to an operator $\mathcal{O}^{global}_U(g)$ supported only in a region $U\subset \partial X$ by restricting the domain of integration. This is the analog of the restricted global boundary symmetry operator considered in the SPT case above and in \cite{EN}.

The difference between the case where we consider $G_{global}$ as part of a higher symmetry as in the previous section and the case at hand where $G_{global}$ is an ordinary symmetry is that this restricted global boundary symmetry operator is not gauge invariant. Indeed,
\[\mathcal{O}^{gauge}_U(f)\mathcal{O}^{global}_U(g) = \exp\big(2\pi i\int_U \tilde\Omega_2(\phi,g)-\frac{1}{3}(a+\delta f)\beta_2(\phi,g)+\frac{1}{3}f\beta_1(\phi)\big)\]
\[\mathcal{O}^{global}_U(g)\mathcal{O}^{gauge}_U(f) = \exp\big(2\pi i\int_U \tilde\Omega_2(\phi,g)-\frac{1}{3}a\beta_2(\phi,g)+\frac{1}{3}f\beta_1(\phi+g)\big).\]
To equate these, we use $\beta_1(\phi+g) = \beta_1(\phi)+\delta \beta_2(\phi,g)$, but then we have to integrate by parts and we get a boundary term
\[\mathcal{O}^{gauge}_U(f)^{-1}\mathcal{O}^{global}_U(g)\mathcal{O}^{gauge}_U(f)=\mathcal{O}^{global}_U(g)\exp\big(2\pi i\int_{\partial U} f\beta_2(\phi,g)\big).\]
Thus, it is impossible to restrict the global symmetry action to a subregion of the boundary in any gauge-invariant way. This implies that it is impossible to realize the $G_{global}$ symmetry action as an ordinary symmetry as a finite depth quantum circuit in the anomalous theory. The global symmetry is fundamentally long-range. We propose this as the salient microscopic feature of anomalies which cannot be cancelled by anomaly in-flow. Recall that while this system lives at the boundary of an SET with all symmetries preserved, it does not satisfy anomaly in-flow because one cannot eliminate the $G_{gauge}$ field $a$ from the bulk; one cannot realize it symmetrically on the boundary of a short-range-entangled phase.

\subsection{Mollifying Nonlocality with Higher Symmetry}

When we promote $G_{global}$ to a 2-symmetry, we now have $\hat\lambda$, which is a local operator, to play with. If we make the shift
\[\mathcal{O}^{global}_U(g) \mapsto \exp\big(-\frac{2\pi i}{3} \int_U a \hat \lambda\big) \mathcal{O}^{global}_U(g)\]
we can fix this. We compute
\[\mathcal{O}^{gauge}_U(f)^{-1}\exp\big(-\frac{2\pi i}{3} \int_U a \hat \lambda\big)\mathcal{O}^{global}_U(g)\mathcal{O}^{gauge}_U(f) \]\[= \exp\big(-\frac{2\pi i}{3} \int_U a \hat \lambda\big)\exp\big(-\frac{2\pi i}{3} \int_{\partial U} f \hat \lambda\big)\mathcal{O}^{global}_U(g)\exp\big(2\pi i\int_{\partial U} f\beta_2(\phi,g)\big).\]
Now when we move the global transformation operator past $\hat\lambda$, $\lambda$ transforms by $\beta_2(\phi,g)$, exactly cancelling the phase on the right hand side. It was critical that we had an edge variable transforming this way and this is precisely what we get when $G_{global}$ is upgraded to a 2-symmetry. Thus, we were able to make the restricted global symmetry operator gauge invariant and indeed from \cite{KT2} we know that such a global 2-symmetry can be implemented by a finite depth quantum circuit.

\section{Decorated Walker-Wang Model for 2-SPT Bulk}
Now we decorate a Walker-Wang model so that it realizes the $3+1 d$ bulk \eqnref{4daction}. We begin with a field theoretic description of the Walker-Wang model, and decorate the strings of the Walker-Wang model with a certain $1+1d$ action. Throughout this section we restrict our attention to $\mathbb{Z}^{(k)}/{p^2}$ models, with $\text{gcd}(2k,p)=1$.\footnote{In the notation of \cite{PBthesis}, this is Walker-Wang model based on tensor category $\mathbb{Z}^{(k)}_{p^2}$. See also \cite{CVK1,CVK2}.}

The $\mathbb{Z}^{(k)}/{p^2}$ Walker-Wang model is described by a two form $H \in \{0,1,\ldots,p^2-1\}$ with Lagrangian density

$$
\mathcal{L}_{\text{WW}}=-\frac{k}{p^2} H \cup H, 
$$

along with the condition $\delta H=0 \mod p^2$. To decorate this Walker-Wang model, and obtain obtain \eqnref{4daction}, start with $\mathcal{L}_{\text{WW}}$ and extend the Hilbert space so as to include 2-form $B$ and flat $\mathbb{Z}/p\times \mathbb{Z}/p$ gauge field $A=(A_1,A_2)$. Consider the Lagrangian

\begin{equation}\label{eq:decWW}
\mathcal{L}_{\text{dec. WW}}=-\frac{k}{p^2} H \cup H + \frac{1}{p^2} H\cup(2k c(A) - p B ). 
\end{equation}
We call this a `decorated' Walker-Wang model. Why? First note $H$ encodes the world-sheets swept out by the Walker-Wang strings (see \cite{CVK2}). Note further that the second term in the action imbues a world-sheet of label $m \in \{0,1,\ldots, p^2-1\}$ (determined by $H\mod p$) with an  action $\omega_m:=\frac{m}{p^2}\left(2k c(A) - p B\right)\mod 1$. Note that, if $m=0\mod p$, the world-sheet action simplifies to $\frac{m}{p^2 2k} c(A)$, which is a $H^2(B\mathbb{Z}/p\times \mathbb{Z}/p, \RR/\ZZ)$ Dijkgraaf-Witten action. When $m\neq 0\mod p$, the action is more complicated, and does not define a legitimate worldsheet TQFT -- we will comment on this in a moment

Let $C = c-p xB$, where $2kx = 1 \mod p$. Then $C$ is a closed $\ZZ/p^2$ 2-form so we can shift $H$ to rewrite \eqnref{eq:decWW} as
$$
\mathcal{L}_{\text{dec. WW}}=-\frac{k}{p^2} (H+C)\cup (H+C)  + \frac{k}{p^2} C\cup C.
$$
We can then expand the second term to obtain 
$$
-\frac{k}{p^2} H\cup H + \frac{k}{p^2} C\cup C,
$$
modulo integers. The first term is the Walker-Wang Lagrangian density, while the second term yields the 3D bulk action \eqnref{4daction}. Hence, by appropriately decorating the Walker-Wang model \eqnref{eq:decWW}, and integrating out the Walker-Wang degrees of freedom $H$, we are left with the bulk action \eqnref{4daction}.

As mentioned above, the 2d action
\[\frac{2k}{p^2}\int c(A) - xp B\]
we use to decorate the Walker-Wang sheets does not define a legitimate 2d TQFT. One sees this by computing the partition function on a 2-sphere and seeing it's zero. Equivalently, the ungauged version of this anomalous 2d TQFT does not have a symmetric ground states. Indeed, the SPT ground state would be of the form
\[\sum_{\phi,\lambda} \exp\big(\frac{2k}{p^2} \int_{S^1} c_1(\phi) - xp \lambda \big)|\phi,\lambda\rangle\]
but this state has charge -1 under the 1-form symmetry. The ground states of this decorated model live in a Hilbert space with states of form $\left|H,\phi,\lambda\right>$. For fixed $H$ we have just shown that the states are not 1-form symmetric. On the other hand, from \secref{2sptbulk} we know that the 2-SPT ground states are 1-form symmetric. Hence summing out the $H$ degrees of freedom has the effect of restoring 1-form symmetry.\footnote{A 1+1d example is the SPT phase $\frac{1}{p}A_1 \cup A_2$ that generates $H^2(B\ZZ/p\times\ZZ/p,U(1))$. One can think of this as a phase where $A_1$ domain walls are labelled with a 0+1d "SPT" for $A_2$. The latter phase doesn't have an invariant ground state: it's simply a particle with a unit of $A_2$ charge. So this decoration just means that the $A_1$ domain wall carries a unit of $A_2$ charge.}





We end this section with a remark about a similar procedure that was performed in \cite{ChenFidkowski} to produce an SPT bulk state from a decorated Walker-Wang model. We do so because our reasoning above gives a direct way of seeing why their construction works. The starting point in the 3D semion Walker-Wang model
$$
-\frac{1}{4} H \cup_{\text{pont}} H
$$
where $H$ is a $\mathbb{Z}_2$ closed 2-form, and $\cup_{\text{pont}}$ is a slightly more complicated version of a cup product, called the `Pontryagin square' \cite{KT2}. They then decorate the string worldsheets with a $1+1d$ $\mathbb{Z}/2\times \mathbb{Z}/2$ topological action furnished by cocycle $c= A_1 \cup A_2$, where $A=(A_1,A_2)$ is a  $\mathbb{Z}_2\times \mathbb{Z}_2$ gauge field. This means that the action above is modified to
$$
-\frac{1}{4} H \cup_{\text{pont}} H+ \frac{1}{2} H c(A).
$$
As above, we can complete the square and integrate out $H$, obtaining
$$
\frac{1}{4} c \cup_{\text{pont}} c
$$
which generates $H^4(B\mathbb{Z}_2 \times \mathbb{Z}_2, \mathbb{R}/\mathbb{Z})$ \cite{KT2}.

\section{Dijkgraaf-Witten Theory in $d$ Dimensions}\label{DWgeneral}

A Dijkgraaf-Witten theory with gauge group $G_{gauge}$ in $d$ spacetime dimensions has a gauge field $a:X \to BG_{gauge}$ and is determined by a cohomology class $\omega \in H^d(BG_{gauge},U(1))$. The action can be written
$$
\int_X a^*\omega \in U(1).
$$
A $G_{global}$ symmetry of this theory is the data of an action of $G_{global}$ on $G_{gauge}$ leaving $\omega$ invariant (up to exact terms) and an extension
$$
G_{gauge} \to G_{total} \to G_{global}
$$
respecting this action. Gauging the symmetry means coupling the theory to a flat $G_{global}$ gauge field $A$. The $G_{global}$ gauge field and the $G_{gauge}$ gauge field $a$ combine to form a $G_{total}$ gauge field $\widehat A:X \to BG_{total}$. In turn, the coupling means that we must extend $\omega$ to a cohomology class $\widehat\omega \in H^d(BG_{total},U(1))$. This is not always possible. In such a situation the $G_{global}$ symmetry is anomalous.

How can we detect the anomaly? We attempt to construct $\widehat\omega$ directly, proceeding order by order in the $G_{global}$ gauge field $A$. The naive extension, $\omega_0(A,a) = \omega(a)$ does not depend on $A$, but may still fail to be gauge invariant because $G_{global}$ acts on $G_{gauge}$. We compute
$$
\delta \omega_0(A,a) = A^*\kappa_1(a) + O(A^2),
$$
where $\kappa_1(a) \in H^1(BG_{global},H^d(BG_{gauge},U(1)))$ is the leading obstruction, and $O(A^2)$ denotes terms in $H^{\geq 2}(BG_{global},H^*(BG_{gauge},U(1)))$. We can cancel the leading obstruction by considering modified action $\omega_1 = \omega_0 - A^*\alpha_1(a)$ where $\alpha_1$ is a 0-cochain on $BG_{global}$ with values in $H^{d-1}(BG_{gauge},U(1))$ which satisfies $\delta \alpha_1 = \kappa_1$. The term $A^*\kappa_1$ can be thought of as the first-order variation of the action under a local symmetry transformation, so it should be thought of as the Noether current for this symmetry (though it is discrete).

The existence of $\alpha_1$ is equivalent to the cohomology class $[\omega]$ being $G_{global}$-invariant. Indeed, one can consider the prism $BG_{gauge} \times [0,1] \subset BG_{total}$ lying over the edge $[0,1] \subset BG_{global}$ labeled by $g \in G_{global}$. If we integrate $\delta \omega_0$ on the length of this prism, we get a form on the boundary equal to $\omega_0(a) - g\omega_0(a)$. By assumption that the action $\int \omega_0(a)$ is $G_{global}$ invariant, there is an $\alpha(a,g)$ satisfying $\omega_0(a) - g \omega_0(a) = \delta\alpha(a,g)$. By assigning to the edge $g$ the form $\alpha(a,g)$ on $BG_{gauge}$, we define $\alpha_1$. This is a discrete version of Noether's theorem.

Next we compute
$$
\delta \omega_1(A,a) = A^*\kappa_2(a) + O(A^3)
$$
where $\kappa_2 \in H^2(BG_{global},H^{d-1}(BG_{gauge},U(1)))$ is the first possibly non-trivial obstruction. If $\kappa_2$ is not exact in this cohomology group, then there is an anomaly. Otherwise, we may define $\alpha_2$ as above and proceed to the next step.

At the end, if all the $\kappa$'s vanish, we will have constructed a closed extension $\widehat\omega$ of $\omega$ and will have successfully gauged the $G_{global}$ symmetry. If any one of $\kappa_2, \kappa_3, ..., \kappa_{d+1}$ are non-vanishing, there is an anomaly. In the mathematics literature, this computational idea is called the Serre spectral sequence and the $\kappa$'s are the differentials. Note that $\kappa_j$ has to vanish to define $\kappa_{j+1}$ and $\kappa_{j+1}$ depends on how we trivialize $\kappa_j$.

Suppose all but the last $\kappa_{d+1} \in H^{d+1}(BG_{global}, H^0(BG_{gauge}, U(1)))$ vanish. In other words,
$$
\delta \omega_d = A^*\kappa_{d+1}.
$$
The above formula means that we can put the gauged theory with action
$$
\int \omega_d(A,a)
$$
on the boundary of a Dijkgraaf-Witten theory with gauge group $G_{global}$ and action $\int\kappa_{d+1}(A)$ (which does not depend on $a$) and have anomaly in-flow cancellation. This is the situation where our anomaly lives on the boundary of an SPT phase.

If $\kappa_k$ for $k<d+1$ is non-vanishing, however, we can attempt to cancel the anomaly using a $d+1$-dimensional theory with action $\int\kappa_k(A,a)$, but this action will necessarily depend on the original dynamical field $a$, so it is outside the bounds of normal anomaly in-flow--only the new gauge field is allowed to live in the bulk. In this situation, the anomalous theory may live on the boundary of a symmetry {\it enriched} topological phase.

Suppose we are in the opposite situation of $\kappa_2 \neq 0$, the most severe anomaly. Then $\kappa_2 \in H^2(BG_{global}, H^{d-1}(BG_{gauge},U(1))$ defines a group extension
$$
H^{d-1}(BG_{gauge},U(1)) \to \cG_{global} \to G_{global}.
$$
General facts about cohomology imply that $\kappa_2$ pulled back to $B\cG_{global}$ is exact. Therefore, if we reinterpret our $G_{global}$ symmetry as a $\cG_{global}$ symmetry and gauge $\cG_{global}$ instead, then $\kappa_2$ is trivialized and we may proceed to $\kappa_3$, which will naturally be defined on $H^3(B\cG_{global},H^{d-2}(BG_{gauge},U(1)))$ the same way it was defined before.

An example of this anomaly can be made in $D= 2+1$ spacetime dimensions with $G_{global} = \ZZ/2_X \times \ZZ/2_Y$, $G_{gauge} = \ZZ/2_{Z_1} \times \ZZ/2_{Z_2} \times \ZZ/2_{Z_3}$, where the global symmetries $X$ and $Y$ commute only up to the gauge transformation $Z_1$. This modifies the flatness condition for the $G_{gauge}$ field $(a_1,a_2,a_3)$ to
\[\delta a_1 = A_1 A_2\]
\[\delta a_2 = 0\]
\[\delta a_3 = 0.\]
If we give the $G_{gauge}$ field the cubic topological term
\[\omega_0(a,A) = \frac{1}{2} a_1 a_2 a_3,\]
then we compute
\[\delta\omega_0(a,A) = \frac{1}{2} A_1 A_2 a_2 a_3 = A^*\kappa_2(a),\]
with $\kappa_2$ a nontrivial class in $ H^2(BG_{global},H^2(BG_{gauge},\RR/\ZZ))= H^2(B\ZZ/2 \times \ZZ/2, \ZZ/2^3)$. We can extend the global symmetry group to the Heisenberg group $D_4 = \langle X,Y,Z| X^2=Y^2=Z^2=1, XYX^{-1}Y^{-1} = Z, Z\ {\rm central}\rangle$ where $Z$ acts by gauge transformation $Z_1$. This fixes the anomaly. Indeed, if we extend $G_{global}$, when we gauge this symmetry then besides our original $\ZZ/2\times\ZZ/2$ gauge field $(A_1,A_2)$ there is a new $\ZZ/2$ gauge field $A_3$ corresponding to the generator $Z$ in the Heisenberg group. Because of the relations in the Heisenberg group, this component is not flat on its own but rather it satisfies $\delta A_3 = A_1 A_2$. We can use it to make the modification
\[\omega_2(a,A) = \frac{1}{2} a_1 a_2 a_3 + \frac{1}{2} A_3 a_2 a_3.\]
Now we compute $\delta \omega_2(a,A) = 0$, so $\omega_2$ is a good gauged action. In the notation above, we've used $\alpha_2 =\frac{1}{2} A_3 a_2 a_3.$ 

Even more interesting is the case $\kappa_2 = 0$ but $\kappa_3 \neq 0$. In this situation, $\kappa_3 \in H^3(BG_{global}, H^{d-2}(BG_{gauge},U(1)))$ defines not a group extension, but a {\it 2-group} extension. The resulting 2-group $\cG_{global}$ has $\Pi_1 = G_{global}$, $\Pi_2 = H^{d-2}(BG_{gauge},U(1))$, action of $\Pi_1$ on $\Pi_2$ induced by the action of $G_{global}$ on $G_{gauge}$, and Postnikov invariant $\kappa_3$. By general nonsense \cite{NSS}, $\kappa_3$ pulled back to $B\cG_{global}$ is exact, so as before if we reinterpret our $G_{global}$ symmetry as part of a 2-group symmetry, then $\kappa_3$ is canonically trivialized and we may move on.

This is precisely the situation we have been studying where $d=3$, $G_{global} = \ZZ/n \times \ZZ/n$ and $G_{gauge} = \ZZ/n$, so $H^{d-2}(BG_{gauge},U(1)) = \ZZ/n$ and we have ended up gauging the 2-group with $\Pi_1 = \ZZ/n \times \ZZ/n$ and $\Pi_2 = \ZZ/n$ and Postnikov invariant $\kappa_3 = 2\gamma_c$. For this 2-group, $\kappa_3$ is trivialized, but the final obstruction, $\kappa_4$ is non-vanishing. However, since it is the last obstruction, as in the ordinary group case, the anomaly may be cancelled by inflow by placing the 3d theory on the boundary of a 4d $\cG_{global}$ SPT with group cocycle $\kappa_4 \in H^4(B\cG_{global},U(1))$. Of course, since $\cG_{global}$ is a 2-group, this is a 2-SPT in the sense of \cite{KT2}. This argument generalizes to prove theorem 1 stated in the introduction.

In fact it is also possible to continue this way to completely cancel the anomaly if we let $\cG_{global}$ be a $d$-group. Let us suppose we are back in the situation where the anomaly can be cancelled by an SPT, ie. $\kappa_2 = \kappa_3 = ... = \kappa_d = 0$, $\kappa_{d+1} \neq 0$. Then it is possible to form the $d$-group extension
$$
B^dU(1) \to \cG_{global} \to G_{global}
$$
which has $\Pi_1 = G_{global}$, $\Pi_2 = \Pi_3 = ... = \Pi_{d-1} = 0$, $\Pi_d = U(1)$ and Postnikov invariant $\kappa_{d+1}$. If we reinterpret the $G_{global}$ symmetry as a $\cG_{global}$ symmetry, then there is no anomaly. This may seem strange, but recall that the gauge field for $\cG_{global}$ is composed of a (flat) $G_{global}$ valued 1-form gauge field $A$ and a $U(1)$-valued $d$-form gauge field $B$ satisfying the constraint
$$
\delta B = \kappa_{d+1}(A).
$$
This restricts the configuration of $A$ to one where $\kappa_{d+1}$ is exact, otherwise there are no solutions to the constraint equation\footnote{ This way to fix the anomaly is almost cheating, however: we cannot impose the above constraint with a Lagrange multiplier since it is an equation of $d+1$-forms and our spacetime is only $d$ dimensional. In the other cases, it is possible to introduce the higher form fields as dynamical fields and impose the Postnikov constraints that describe $\cG_{global}$ with Lagrange multipliers or even with potential wells.}. In this situation, the action is
$$
\int \omega_{d}(A) - B,
$$
in other words, all of space (which is compact!) is electrically charged under $B$. This is a typical though under-appreciated feature of 't Hooft anomalies. Our argument shows that all anomalies of Dijkgraaf-Witten theories may be interpreted as a non-zero total charge for a certain higher symmetry.

We have of course ``cheated" by cancelling the anomaly this way. The group $U(1)$ is not discrete, so the path integral requires regulation, perhaps with a kinetic term. We can do a bit better, since for finite $G_{global}$ the anomaly will land in the image of  $H^{d+1}(BG_{global}, \ZZ/n)$ for some $n$. The we are back to the finite path integral we've been using. However, another problem is that no possible d-dimensional action principle can impose the constraint $\delta B = \kappa_{d+1}(A)$ above. Thus, while we have ``cancelled" the anomaly, we have broken the field theory in other ways. Extending to higher symmetry is only able to buy us anomaly in-flow, not freedom from the anomaly.

\section{More General 2+1d TQFT's}\label{gentqft}

In \cite{ENO}, the authors consider the problem of gauging a symmetry $G_{global}$ of a modular tensor category $\cC$, which is supposed to encode the quasiparticle spectrum, fusion, and braiding statistics in a (modular) 2+1d TQFT. A good review for condensed matter physicists and the source of our notation is \cite{BBCW}.

The leading obstruction to gauging is an element $\mathfrak{D}\in H^3(BG,\mathcal{A})$, where $\mathcal{A}$ is the group of abelian anyons in $\cC$. We will argue that this obstruction can be discarded if we gauge $G_{global}$ as part of the 2-group $\cG_{global}$ with $\Pi_1 = G_{global}$, $\Pi_2 = \mathcal{A}$, and $\beta = \mathfrak{D}$. In fact we will find that gauging $G_{global}$ leads to an emergent 1-form $\mathfrak{D}$ symmetry.

The $H^3$ obstruction comes from considering the associativity of the action of $G_{global}$. For three elements $g_1,g_2,g_3 \in  G_{global}$ and $a \in \cC$, define $\Omega_a(g_1,g_2,g_3)$ as the braiding phase depicted below.
\begin{center}
\begin{tikzpicture}
\node (pic) {\includegraphics[width=0.4\textwidth]{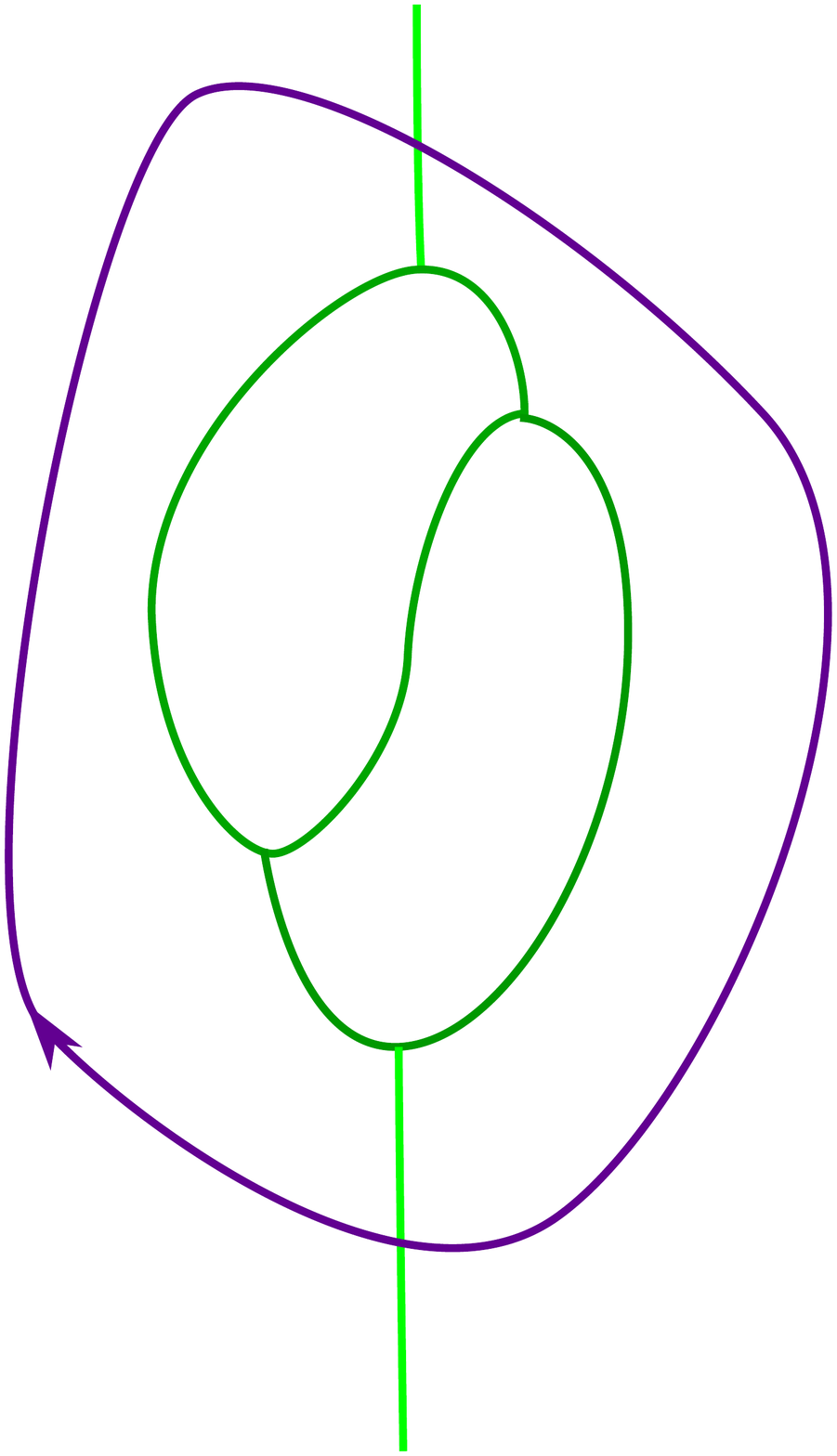}};
\node at (0,4) {$g_1g_2g_3$};
\node at (0.9,2) {$g_2g_3$};
\node at (-1,-1) {$g_1g_2$};
\node at (1.3,0) {$g_3$};
\node at (0,0) {$g_2$};
\node at (-1.6,0.3) {$g_1$};
\node at (-1.6,-2) {$a$};
\node[text width=6cm] at (0,-4.5) {An anyon $a\in\cC$ encircles a codimension 2 defect where three domain walls $g_1, g_2, g_3$ associate.};
\end{tikzpicture}
\end{center}
In our Dijkgraaf-Witten example in \secref{Background}, this phase is the product of the braiding phases around each of the Y-zippers. For $a$ the unit charge Wilson line, we thus compute\footnote{This reproduces formula (102) in \cite{BBCW} in this case with $\beta(g_1,g_2) = \exp(2\pi i c(g_1,g_2)/n)$.}
$$
\Omega(g_1,g_2,g_3) = \exp(2\pi i[c(g_1,g_2g_3)+c(g_2,g_3)-c(g_1g_2,g_3)-c(g_1,g_2)]/n).
$$
Note that the exponent in $\Omega$ is $\delta c$, which is divisible by $n$, so the electric line sees no charge. However, if $a$ is a unit charge magnetic line, we get a phase
\[\Omega(g_1,g_2,g_3) = \exp(-4\pi i[c(g_1,g_2g_3) + c(g_2,g_3)-c(g_1g_2,g_3)-c(g_1,g_2)]/n^2).\]
It is argued in \cite{BBCW} that $\Omega_a(g_1,g_2,g_3)$ can always be expressed as a braiding phase of $a$ with an abelian anyon $\mathfrak{D}(g_1,g_2,g_3)$ independent of $a$. In our Dijkgraaf-Witten example, one computes $\mathfrak{D}(g_1,g_2,g_3)$ as an anyon with electric charge
$$
-2[c(g_1,g_2g_3)+c(g_2,g_3)-c(g_1g_2,g_3)-c(g_1,g_2)]/n^2 = -2\gamma_c(g_1,g_2,g_3).
$$
This is precisely the obstruction we found in field theory  \secref{ss:anomalyinflow}. Indeed, it is shown in \cite{ENO} that something must be done to restore associativity if we are to end up with an honest $G_{global}$ gauge theory. In \cite{BBCW} it explained how this is an obstruction to symmetry fractionalization. If
$$
\mathfrak{D} = \delta \mathfrak{Z},
$$
for some cochain $\mathfrak{Z}(g_1,g_2) \in C^2(BG,\mathcal{A})$, then we can locally modify the $G_{global}$ action, decorating the Y-zippers where $g_1$ and $g_2$ meet with the abelian anyon $\mathfrak{Z}(g_1,g_2)^{-1}$. These will precisely cancel the phase $\Omega_a$ and we will be left with an associative action. After doing this, one may encounter a second obstruction $\Omega(g_1,g_2,g_3,g_3) \in H^4(BG_{global},U(1))$ depending on $\mathfrak{Z}$. This anomaly can be cancelled by in-flow from a $G_{global}$ SPT in 3+1d with cocycle $\Omega$ \cite{CLV}.


If we are unable to cancel $\Omega_a$ in this fashion (as in \secref{ss:anomalyinflow}), or simply choose not to, then when domain walls proliferate so will the codimension 2 associator bubbles considered above. But the associator bubbles behave like anyons in $\mathcal{A}$, so we are proliferating anyons. Proliferating anyons can be thought of as gauging a 1-form symmetry. The resulting theory thus has a global symmetry $G_{global}$ as before, as well as an emergent $1$-form symmetry parameterized by $\mathcal{A}$. Thus if we don't modify the group action and allow invertible anyons to proliferate, we end up with a 2-gauge theory with $\Pi_1 = G_{global}$, $\Pi_2 = \mathcal{A}$ and $\beta = \mathfrak{D}$. Here $\beta$ encodes how the associator bubbles act on $\mathcal{C}$. 

We can rephrase the above statements as follows. The world-lines corresponding to the associator bubbles are Poincar\'e dual to a 2-form B valued in $\mathcal{A}$. In this alternative formalism, closedness of $G_{global}$ domain walls corresponds to a flatness condition for the $G_{global}$ 1-form gauge field $A$:
$$
\delta A = 0.
$$
Meanwhile, the creation of $B$ lines by the formation of associator bubbles of $G_{global}$ domain walls creates a twisted flatness condition for the $\mathcal{A}$ 2-form gauge field $B$:
$$
\delta B = \mathfrak{D}(A).
$$
Together, these form the data of a flat $\cG_{global}$ gauge field. One sees that this second constraint trivializes the $H^3$ obstruction, but to define a consistent path integral, we must integrate over trivializations $B$.

Proceeding, we may encounter one last obstruction, coming from the braiding of two Y-zippers around each other. The calculation in \cite{ENO} shows that when we fix $B$, this obstruction defines a cocycle in $H^4(BG,U(1))$ roughly equal to the Pontryagin square of $c$. It is easy to check that the cocycle is also invariant under 1-form gauge transformations of $B$, so that the obstruction is actually a cocycle on the whole 2-group, $\alpha\in H^4(B\cG_{global},U(1))$. Thus, this last obstruction can be cancelled by placing the gauged theory on the boundary of a $\cG_{global}$ Dijkgraaf-Witten theory with cocycle $\alpha$. The interpretation is that the anomalous gapped phases with $H^3$ anomaly can be realized on the boundary of a 2-SPT. 

\section*{Acknowledgements}
RT would like to thank Anton Kapustin for pointing out the possibility of anomaly in-flow for ordinary global symmetries on the boundary of 2-SPT's. RT is supported by the NSF GRFP grant. CVK is supported by the Princeton Center for Theoretical Sciences, and acknowledges useful conversations with Fiona Burnell and Lukasz Fidkowski.

\section*{Appendix: Cochains, Cocycles, ...}

Consider a 2+1d Dijkgraaf-Witten theory with gauge group $G_{gauge} = \ZZ/3$ and action
$$
S_{3d} = \frac{1}{3} \int \tilde{a} \cup \frac{\delta \tilde a}{3}.
$$

For us, the $\ZZ/3$ gauge field $a$ consists of an integer $a_{ij}$ defined mod 3 for each edge $ij$ in some triangulation of spacetime. This data is called a $\ZZ/3$ valued 1-cochain. Similarly we will make use of more general $G$ valued $p$-cochains, which are the data of an element of $G$ for every $p$-dimensional face. The continuum analogue of this action is an abelian Chern-Simons theory with bilinear form
$$
K=
\begin{pmatrix}2 & 3\\
3 & 0
\end{pmatrix}.
$$

For each oriented triangle $i<j<k$, the gauge field $a$ satisfies a flatness condition
\[ a_{ij} + a_{jk} = a_{ik} \mod 3\]
making it a $\ZZ/3$ 1-cocycle. There are also $G_{gauge}$ transformations, each parametrized by an element $f_i \in \ZZ/3$ for each vertex $i$ and acting by
\[ a_{ij} \mapsto a_{ij} + \delta f_{ij} = a_{ij} + f_i - f_j \mod 3.\]
The gauge equivalence class $[a]$ is thus a $\ZZ/3$ degree 1 cohomology class, ie. $[a] \in H^1(X,\ZZ/3)$, where $X$ denotes the 3d spacetime. 

The quantity $a_{ij}$ should only be referred to in equations defined mod 3. We will need something more flexible to be explicit with the formulas we manipulate. The notation $\tilde a$ will be used more generally and means an integer $\tilde{a}_{ij} \in \{0,1,2\}$ equal to $a_{ij}$ mod 3 for each edge $ij$. In particular, this means it satisfies
\[ \tilde{a}_{ij} + \tilde{a}_{jk} = \tilde{a}_{ik} \mod 3\]
and transforms under the $G_{gauge}$ transformations. The quantity $\tilde{a}$ is useful for expressing things like the Bockstein
\[ Bock(a) := \frac{\delta \tilde{a}}{3}.\]
Indeed, $\delta \tilde{a}_{ijk} = \tilde{a}_{ij} + \tilde{a}_{jk} - \tilde{a}_{ik}$ is not necessarily zero, only divisible by three, so the expression above makes sense as a 2-cochain and is even closed (but not necessarily exact!).

All the expressions we deal with will be functions truly only of $a$, not its integer lift $\tilde a$. This property could be thought of as compactness of the gauge group, analogous to $G=U(1)$ vs. $G=\RR$. For example, we could make a different convention for producing the integer lift $\tilde a$, and this would amount to a shift
\[ \tilde a \mapsto \tilde a + 3x,\]
for any arbitrary integer 1-cochain $x$. If we perform this shift on the Bockstein, we get
\[ Bock(a) \mapsto Bock(a) + \delta x,\]
so as a cohomology class, $Bock(a)$ only depends on $a$, not its integer lift $\tilde a$. The Dijkgraaf-Witten action itself transforms into
\[ S_{3d} \mapsto S_{3d} + \int x \cup Bock(a) + \frac{1}{3} \int \tilde a \cup \delta x + \int x \cup \delta x.\]
The action appears in the path integral as $\exp(2\pi i S)$, so it really only matters modulo integers. Indeed, the first and third variations above are integers, so they don't contribute to any observables. Meanwhile, the second variation can be integrated by parts to obtain
\[ \frac{1}{3} \int \delta \tilde a \cup x,\]
which is also an integer since $\delta \tilde a$ is divisible by 3.

The action $S_{3d}$ should also only depend on the gauge equivalence class $[a]$. To check, we shift $a\mapsto a + \delta f$ and see that we get the same thing. Let's start with the Bockstein.
\[Bock(a) \mapsto \frac{1}{3} \delta( \widetilde{a + \delta f}).\]
It's important to be careful about the ordering of tildes and deltas. Indeed, while $\delta^2 = 0$, $\delta( \widetilde{\delta f}) \neq 0$. However, we can still find an integer valued 1-cochain $Bock_1(a;f)$ satisfying
\[ Bock(a+\delta f) - Bock(a) = \delta Bock_1(a;f).\]
This is called the first descendant of the Bockstein. Descendants of various cohomology operations will appear repeatedly in this paper. We don't compute $Bock_1(a;f)$ right now. Instead we compute $Bock_1(0;f)$. Note the difference between $\widetilde{x+y}$ and $\tilde{x}+\tilde{y}$ is because of carrying. That is,
\[ \widetilde{f_i - f_j} = \tilde{f_i} - \tilde{f_j} + 3(f_j > f_i),\]
where $(f_j > f_i)$ is 1 if $\tilde{f_j} > \tilde{f_i}$, 0 otherwise. One could think of this in quantum information terms as a generalized ``control-Z" gate. Using this we compute
\[ \frac{1}{3} \delta (\widetilde{\delta f})_{ijk} = (f_j > f_i) + (f_k > f_j) - (f_k > f_i).\]
Thus we can take
\[ Bock_1(0;f)_{ij} = (f_j > f_i).\]
The complete $Bock_1(a;f)$ is some version of this where the comparison $f_j > f_i$ is made ``after shifting zero by $a_{ij}$". We leave its computation as an exercise to the interested reader.

Anyway, given such a quantity, the action transforms by
\[ S_{3d} \mapsto S_{3d} + \int Bock_1(0;f) \frac{\delta \tilde a}{3} + \frac{1}{3} \int \tilde a \delta Bock_1(a;f) + \int Bock_1(0;f) \delta Bock_1(a;f).\]
The first and third variations are integers, while the second may be integrated by parts to give an integer. Therefore, the action $S_{3d}$ is gauge invariant.

\bibliographystyle{unsrt}

\end{document}